\documentclass{llncs}

\usepackage{vdm-rg}
\usepackage{xfrac}
\leftRecord
\leftCases
\usepackage{graphicx}

\usepackage{textcomp}

\usepackage{xcolor}

\usepackage{hyperref}
\usepackage{enumitem}

\newtheorem{lemm}[theorem]{Lemma}

\newcommand{\redText}[1]{\color{red}#1\color{black}}

\newcommand{\defeq}{\stackrel{def}{=}}

\begin{document}


\title{Extending Rely-Guarantee thinking to handle
Real-Time Scheduling}


\author{Cliff B Jones\inst{1} \and
Alan Burns\inst{2}}
\institute{School of Computing, Newcastle University, UK \and
Department of Computer Science, University of York, UK}
\begin{verbatim}
@article{JonesBurns-23b,
	author = {Jones, Cliff B. and Burns, Alan},
	doi = {10.1007/s10703-023-00441-y},
	journal = {Formal Methods in System Design},
	title = {Extending rely-guarantee thinking to handle real-time scheduling},
	url = {https://link.springer.com/article/10.1007/s10703-023-00441-y},
	year = {2023}}
\end{verbatim}
   
\maketitle
     
\bibliographystyle{alpha}     

\nocite{RandellEt04}
\nocite{BowenEt-23}

\noindent 
\makebox[\linewidth]{\today} \\


\begin{abstract}
The reference point for developing any artefact is its specification;
to develop software formally, a formal specification is required.
For sequential programs, 
pre and post conditions
(together with abstract objects)
suffice;
rely and guarantee conditions extend the scope of formal development approaches to tackle concurrency.
In addition,
real-time systems need ways of both requiring progress and relating that progress to some notion of time.
This paper extends rely-guarantee ideas to cope with specifications of 
--and assumptions about--
real-time schedulers.
Furthermore it
shows how the approach helps identify and specify fault-tolerance aspects of such schedulers by systematically challenging the assumptions.
\end{abstract}


\vfill
{\color{blue}
This article was published (2023-11-30) in {\em Formal Methods in System Design} Please cite as:\\
\begin{verbatim}
@article{JonesBurns-23b,
	author = {Jones, Cliff B. and Burns, Alan},
	doi = {10.1007/s10703-023-00441-y},
	journal = {Formal Methods in System Design},
	title = {Extending rely-guarantee thinking to handle real-time scheduling},
	url = {https://link.springer.com/article/10.1007/s10703-023-00441-y},
	year = {2023}}
\end{verbatim}
}
     
%


\newpage
\section{Introduction}
Rely and guarantee conditions were originally conceived~\cite{Jones81d} as a way to achieve compositional design of shared-variable concurrent programs.
It was subsequently shown~\cite{HayesJacksonJones03,BurnsHayesJones-19} that rely conditions could be used to describe interfaces to external components.
The current paper extends this work to address systems that have real-time constraints -- specifically to
address specifications that incorporate hard deadlines.
In particular,
scheduling requires specifying behaviour with respect to time in the world outside the computer
and the meeting of deadlines requires rigid
progress conditions for the scheduler to be specified.

Informally, scheduling must ensure that jobs will execute so that deadlines are met.
This overall objective can be divided into a planning phase and execution of run-time scheduler software that 
dynamically allocates resources.
Given the extensive literature on scheduling, ``planning'' is almost always going to amount to selecting from a repertoire of known scheduling algorithms.

Crucial inputs to planning are {\bf assumptions} about arrival patterns and estimates of the resources required by each job of the application.
Only if these assumptions are met can the run-time system be expected to adhere to the required deadlines;
the assumptions thus underlie run-time execution
as well as informing the decisions made in planning.

In the non fault-tolerant case, 
most of the real work is undertaken during planning; there are established algorithms
with proven properties.
This paper formalises the ideas for the special case that the only resource to be considered is processor time
and the chosen scheduling approach is EDF (Earliest Deadline First).

A fault-tolerant system is one that can tolerate, to some degree, violation of assumptions.
But weaker assumptions must be articulated and recorded. 
The requirements on the run-time scheduler for
{\bf fault-tolerant behaviours} are significantly more complicated than in the non fault-tolerant case;
their specification must be precise about the assumptions behind the various degraded levels of behaviour.

{\bf Mixed-Criticality Scheduling}~\cite{Ves2007} enhances the robustness of schedulers 
by distinguishing different ``criticality'' levels for classes of  jobs. 
Since Vestal's~\cite{Ves2007} paper in 2007, 
a  wide range of protocols have been proposed and published~\cite{burns:survey:2017}.
The aim of this body of work is to improve the survivability of systems by
providing a variety of degraded behaviours that can take effect if the system experiences overrunning execution
times. Unfortunately the models developed in this literature are usually not formally or precisely defined;
they tend to focus on the algorithmic properties of protocols and provide,
at best, informal descriptions of the actual required run-time behaviour of the defined scheduler.
This paper extends the Rely-Guarantee approach to cover the temporal properties of fault-tolerant,
real-time, Mixed-Criticality Systems (MCSs). 
From this foundation the rigorous implementation of the necessary run-time schedulers can be developed.


\subsection{Contribution}

Two earlier papers~\cite{BurnsJones:ECRTS:2022,JonesBurns-22} explored the basic approach of using rely and guarantee conditions to specify the
run-time behaviour of a scheduler for a simple fixed priority-based real-time MCS. 
Here, 
this framework is extended to address some key elements that were missing from these initial studies.
In particular:

\begin{itemize}
\item Both pre-run-time planning and run-time scheduling are addressed and their relationship defined (see Section \ref{S-plan-sched}).
\item Type invariants and relations are used to capture assumptions (see Section \ref{S-State}).
\item External and internal definitions of time are employed together with a notion of precision to formalise their relation (see Section \ref{S-rho}).
\item The deadlines of jobs and the passage of (external) time are used to require progress
(this removes the need for the practice of augmenting rely-guarantee conditions with some form of temporal logic) (see Section \ref{S-time}).
\item The EDF scheduling scheme is employed (see Section \ref{S-EDF-spec}).
\item Fault-tolerance is addressed by systematically challenging these invariants and assumptions --
and characterising weaker ones where appropriate (see Section \ref{S-FaultT}).
\end{itemize}

\subsection{Plan of paper}


Section~\ref{S-RG} describes the background work on rely and guarantee conditions;
Section~\ref{S-plan-sched} outlines basic aspects of planning and scheduling without considering fault tolerance.
Section~\ref{S-f-Scheduler} sets about formalising a specification for the general case and 
 a specific implementation
(EDF) 
is the focus in Section~\ref{S-EDF-spec}.
The objective is to show how the chosen approach to formalism can both identify aspects of robustness and be used to
characterise fault-tolerance:
this is the subject of Section~\ref{S-FaultT}.
The paper concludes with a section that summarises the results and relates them to other publications.

\section{The Rely Guarantee Framework}
\label{S-RG}

%


Tony Hoare's~\cite{Hoare69a} paper is central to much of the work on program verification;
moreover~\cite{Hoare71a} sets the scene for (formal) support of the design process itself.
In VDM (Vienna Development Method)~\cite{Jones80a},
a move is made to post conditions that are relations between the initial and final states of the specified component.
Although this requires slightly more complicated inference rules
(see~\cite{Aczel82,Jones86a}),
relations are required for most non-trivial specifications %
and this approach is used in VDM as well as,
for example,
Z~\cite{Hayes87a,woodcock1996using},
B~\cite{Abrial96},
Event-B~\cite{Abrial10} and
Alloy~\cite{Jackson-12}.

In all of the predicates that are relations
(post, rely and guarantee conditions),
undecorated identifiers refer to the earlier state and primed identifiers apply to the later state,
for example:

\begin{formula}
post-Sort(s, s') \defeq is-ordered (s') \And is-permutation(s', s)
\end{formula}

Pre and post conditions provide sufficient specifications for sequential programs.
Based on such a specification,
a step of development that proposes an implementation consisting of components that will
--for example-- execute sequentially
can be justified solely in terms of the specifications of those sub-components.
Subsequently,
the sub-components can be developed with respect to their specifications and
their designer(s) need not be concerned with either their context or sibling components.
The same can be said for decompositions that employ any of the sequential programming constructs for which inference rules have been recorded.
This ``posit and prove'' development idea affords clear documentation even if the developer has to backtrack.
Technically,
each of the programming language constructs can be shown to be monotone with respect to an ordering where 
a component can be substituted that has a weaker pre condition and/or a stronger post condition.
This notion of compositionality is relatively straightforward for sequential programs.
Furthermore as is shown in~\cite{Jones80a,Jones86a}, 
the posit-and-prove approach extends to the use of data abstraction and reification.


Achieving a suitable notion of compositionality for concurrent programming constructs was bound to be challenging precisely because
of the interference that is at the heart of concurrency.
A pre condition indicates potential starting states but says nothing about the interference that a developed implementation must tolerate;
a post condition relates acceptable final states to their initial counterparts but puts no limit on the intermediate state transitions.

With hindsight,
it is obvious that the specification of a component that experiences interference must include a specification of that interference.
This is precisely what the Rely-Guarantee approach does:
\cite{Jones81d}%
\footnote{Shorter,
and more accessible descriptions include~\cite{Jones83a,Jones83b}.
An excellent discussion of compositionality is contained in~\cite{DeRoever01}.
There is an extensive literature on Rely-Guarantee ``thinking'' ---
see~\cite{HJM-23} for references 
and~\cite{HayesJones18} for a description of an algebraic presentation of Rely-Guarantee ideas.
Furthermore,
combinations with ``Concurrent Separation Logic'' are presented in~\cite{Vafeiadis07,VafeiadisParkinson07,FengSAGL-07}.}
extends pre/post specifications with two extra predicates which are pictured in Figure~\ref{F-RG} %
(where $\sigma\sb{0}$ is a starting state of a computation whose subsequent state transitions extend to the right,
the transition from $\sigma\sb{i}$ to $\sigma\sb{i + 1}$ is an example of a transition by the environment,
in contrast the transition from $\sigma\sb{j}$ to $\sigma\sb{j + 1}$ is an example of a transition by the component
(there could be many transitions of either sort during the execution of a component)
and $\sigma\sb{f}$ is a final state).

\begin{figure}
\begin{center}
$
\underbrace{\overbrace{\sigma\sb{0}}^{\color{blue}pre\color{black}} \hspace{1em} \cdots \hspace{1em} \overbrace{\sigma\sb{i} \; \sigma\sb{i+1}}^{\color{blue}rely\color{black}} \hspace{1em} \cdots \hspace{2em} \underbrace{\sigma\sb{j} \; \sigma\sb{j+1}}\sb{\color{red}guar\color{black}} \hspace{1em} \cdots \hspace{1em} \sigma\sb{f}}\sb{\color{red}post\color{black}}
$
\end{center}

{\color{blue} $pre/rely$ are assumptions the developer is invited to make}

{\color{red} $guar/post$ are commitments that the code must achieve}

\caption{The four predicates forming a rely-guarantee specification}
\label{F-RG}

\end{figure}



{\bf Rely conditions} are relations and can be usefully thought of as post conditions of interfering steps.
Thinking of them in this way makes clear that they must be relations.
Typical examples include that some variable changes monotonically
(integers are trivial examples;
employing sets as data abstractions suggests more interesting examples such as subset ordering being preserved).

It is important to remember that pre conditions provide assumptions that developers can make:
they are invited to assume that their implementation will only have to achieve its post condition
when executed in a starting state that satisfies the pre condition.
In exactly the same vein,
a developer is invited to assume that 
any state transition by the environment will be within the relational rely relation of the component.
In both cases,
the onus is on any deployment decision to ensure that the environment satisfies the assumptions.
On many occasions, 
such a deployment decision will involve proving that employing a piece of specified code in a given context will ensure that the assumptions are satisfied.
This is again the essence of a posit-and-prove approach.
Furthermore, 
it is for this reason that specifications of components include {\bf guarantee conditions} which state the maximum
interference that an implementation can inflict on its environment.
As such,
it is obvious that guarantee conditions must also be relations.
Because interference could make several steps 
(or none),
rely and guarantee conditions are required to be transitive and reflexive.

Although conceived as a decomposition approach for the development of concurrent programs,
it was realised that rely conditions could also be used to characterise the physical environments in which software could be deployed.
Going further~\cite{HayesJacksonJones03,BurnsHayesJones-19} showed how the specification of the way 
in which a cyber-physical system should perform in a physical environment could be used to derive the specification 
of its software control system.
The chosen rely conditions on the physical components need to be reviewed and approved by those making the decision to deploy a system
in a particular physical context.

The term {\bf obligations} is used in this paper to emphasise that there are issues in addition to 
rely and guarantee conditions.
In particular, the preservation of {\bf state invariants} that connect with the progress of time 
(see Section~\ref{S-time} below)
are important. 
Furthermore,
the picture in Figure~\ref{F-RG} needs to be understood in the presence of systems executing with real-time obligations
(again see Section~\ref{S-time}).

Formal rules for proving the validity of a decomposition of a component into concurrent threads 
(whilst satisfying its obligations)
are not needed in the current paper;
all that is used is the intuitive idea  that concurrent threads can co-exist providing their 
rely and guarantee relations match in the sense that 
--for each component-- 
the combined guarantee conditions of any environment components must stay within the rely condition of the specified component.
In what is undertaken below for MCS,
rely and guarantee conditions pinpoint assumptions and are used to check that those on scheduling algorithms 
match those on the jobs that are to be scheduled.

The limited use of formal proofs fits with a message of
``Rely-Guarantee thinking''
which might be trivialised to just saying 
``think about and record assumptions''.
But the fact that there are formal rules underneath the simplified message,
means that formal proofs can be added when considered necessary.

Of further relevance to the current paper is the fact that 
earlier applications led to the use of nested obligations where different sets of rely and guarantee conditions describe alternative behaviours.
In fault-tolerant applications, 
strong rely conditions require optimal behaviour characterised by strong guarantee conditions;
weaker assumptions describe conditions that might have to be tolerated when faults occur but these are paired with weaker
--but still useful-- 
commitments for fault-tolerant behaviour.
As is shown below in Section~\ref{S-FaultT},
changing modes needs careful thought.


The original rely-guarantee publications did not incorporate progress assertions which can state that something will eventually happen;
these were addressed in~\cite{Stolen90,stolen1991attempt,Xu92} and are the subject of current research by Ian Hayes and colleagues
(see, for example,~\cite{HayesEt-18}).
The obvious way to tackle so-called ``liveness'' conditions might be thought to be ``Temporal Logic''
(for an approachable description with good source references see~\cite{schneider97concurrent}).
Despite its name,
temporal logic does not really handle time;
its modal operators facilitate specifying orders of events.
A novel approach to both time and liveness is outlined in~\cite{JonesBurns-22} and is expanded on below in Section~\ref{S-time}.


%
%
%

\section{Planning and scheduling}
\label{S-plan-sched}

The entities that consume resource at run time are here referred to as $Job$s.
Real-Time Scheduling normally categorises jobs into tasks.
(See~\cite{BurnsWellings-16} for more discussion of tasks.)
Tasks do not terminate and reflect the fact that classes of jobs recur.
In the current paper,
jobs are assumed to have types 
(see Sect.~\ref{S-f-planning})
and tasks can be thought of as handling all jobs of a certain type.

A pre-run-time planning phase is concerned with determining the feasibility of both the resource demands of jobs
and their inter-job arrival gap.
Planning is also concerned with designing (or choosing) a
run-time scheduling algorithm that controls the execution of jobs.

Time progresses both during planning and run-time execution but it is useful to employ different granularities 
for discussing the distinct phases;
research on ``timebands''~\cite{BurnsHayes09} has identified the disadvantages of trying to address 
disparate granularities on a single time axis.
In a real-time system, 
jobs have deadlines that must be guaranteed to be satisfied.
A run-time scheduler and the jobs to which it is allocating resource are
items of software that execute within an environment in which the passage of time is explicitly
linked to the progress that the jobs are making towards their deadlines. 

This leads to the following framework for the planning and scheduling of a real-time system:

\begin{formula}
Passage-of-Time \; \parallel \;  \R
Planning\; ; \; \{Scheduler  \; \parallel \;  Job\sb{1}  \; \parallel \;  Job\sb{2}  \; \parallel \; \cdots \; \parallel \;  Job\sb{k} \}
\end{formula}

Planning is a human-centred activity that may be subject to timing constraints measured in days, weeks or even months;
this process is important, but not the focus of this paper. 
Once planning has determined that jobs which fit expectations can be feasibly handled by a chosen scheduling algorithm,
then a $Scheduler$ embodying that algorithm can be used to control a collection of actual $Job$s.
Here the passage of time must be considered at a dramatically finer grain, 
perhaps at the millisecond level or less. 
Neither the scheduler nor any of the jobs can influence the passage of time. 
As time is progressing;
a scheduler must allocate the resources in the manner assumed during planning so that all jobs will terminate by their deadlines.

\subsection{Formalising $Planning$}
\label{S-f-planning}

In this paper,
each job has a type and it is assumed that a task handles jobs of the same type.
Static information that is assumed in $Planning$
is contained in records of type $JobType$:%
\footnote{Modest use is made below of VDM notation about which there are many books 
(including~\cite{Jones90a})
plus an international standard;
comments are provided for less common notation.
Here the record notation should be obvious;
constructor functions are of type $mk-JobType: Duration \x Duration \x \cdots \to JobType$;
field selection is written in a postfix style:
$jt \in JobType \Implies jt.D \in Duration$.}

\begin{record}{JobType}
D: Duration\\
C: Duration\\
\cdots
\end{record}

\noindent
Where the $D$ field contains the relative deadline of a job 
(the extent of time after the job starts by which it should complete)
and the $C$ field contains the estimate of the maximum resource that will be consumed
(typically estimated ``worst case execution time'').%
\footnote{Further fields relating to fault tolerance are added in Sect.~\ref{S-FaultT}.}
It is important to notice that both of these durations relate to time in the world external to the computer.

In addition to the information about execution of $Job$s,
planning must also be based on assumptions about the arrival times of $Job$s.
These assumptions are recorded as a predicate $\mathcal{A}$ over the entire history of the run-time behaviour of the system
(see $\Sigma$ in Section~\ref{S-rho}).
%
%
This could for example record the minimum gap between jobs for sporadic tasks or the maximum jitter on periodic tasks.

For planning,
$\mathcal{A}$ and $C$ are assumptions 
whereas $D$ is a requirement 
that informs the selection of a run-time scheduling algorithm.
A planner has to allow for task switching on (worst-case) arrival patterns but 
this will be far shorter than the gaps between job arrivals.
Planners also need to be aware that internal computer clocks cannot precisely track time in the external world
but this discrepancy should also be small. 


Criteria under which the assumptions establish schedulability are proven in~\cite{cerqueira2016prosa,PROSA-2}
(see Section~\ref{S-related}).

\section{Formalising $Scheduler$s (without fault-tolerance)}
\label{S-f-Scheduler}

This section reviews general properties required of scheduling algorithms;
Sect.~\ref{S-EDF-spec} focusses on one specific scheduling approach.

Resources are allocated at  run-time by a scheduler;
in the case considered here,
processor time is allocated to $Job$s,
In contrast to their use in reasoning about a ``schedulability test'' during $Planning$,
the assumptions $C$ and $\mathcal{A}$ define the run-time
conditions in which the $Scheduler$ must ensure that jobs complete by their deadlines.


A $Scheduler$ runs for an indeterminate period of time so it is not sensible to record its specification with a post condition.
Similarly,
the run-time code has to achieve its effect only if each job consumes less time than its estimated worst-case execution time;
this is an assumption about something that obtains only during the execution of the scheduler and therefore cannot be 
defined in a pre condition.

The ideas outlined in Section~\ref{S-RG}
on adding rely and guarantee conditions to pre/post condition specifications were conceived as 
a way of providing compositional proof rules for concurrency.
It was later observed that rely conditions could characterise acceptable behaviour of components that
--rather than being developed--
were external to the software being specified.
In the case of cyber-physical systems,
recording assumptions about physical components provided a way of deriving the specifications of required control systems
(see~\cite{BurnsHayesJones-19});
in fact,
that paper improved on~\cite{HayesJacksonJones03}
by employing the timeband ideas from~\cite{BurnsHayes09};
aspects of these ideas are employed below.

Since early publications such as~\cite{Jones80a},
VDM has used post conditions that are predicates of the initial and final states of a computation.
Since rely and guarantee conditions characterise interference,
they are also written as relations.
For example,
$x \leq x'$
would express the fact that the value of a state variable named $x$ should not decrease.
As a rely condition,
this might be an assumption that the developer is invited to make;
as a guarantee condition,
it would be a constraint that must be adhered to by the implementation.
Notice how the role of rely conditions as assumptions that the developer is invited to make mirrors the role of pre conditions.
Similarly,
guarantee conditions play a role similar to post conditions in that they must be respected by code execution. 

The idea that pre and post conditions must be satisfactory for any deployment of a specified component
involves showing that the context is such that invocation will only occur in states satisfying the pre condition
and that any state satisfying the post relation will be satisfactory for the contextual task.
Exactly the same dichotomy applies to rely and guarantee conditions.

\subsection{State of run-time model}
\label{S-State}

Specifications of scheduling algorithms can be expressed in terms of a run-time state.
Operations that cope with jobs arriving and finishing can then be specified as predicates over these states.
The fields of the following records are explained after its syntax is given:

\begin{record}{State}
t: ClockValue\\
active: \mapof{JobId}{JobInfo}\\
run: \Opt{JobId}\\
shared-ents: \mapof{Id}{Val}\\
mode: \set{\const{Normal},\const{Ft}}
\end{record}

\noindent
Scheduling software needs a way of tracking time:
unfortunately, this cannot be a completely precise meter of $Time$ in the world external to the computer
so the $t$ field is shown as containing a $ClockValue$.
The relationship between $t$ and external time is made precise in Sect.~\ref{S-rho}.

Schedulers do not control when $Job$s start --- this might for example happen in response to interrupts.
When a job does start,
a $JobInfo$ record appears in the $active$ map of $State$ associated with the new $JobId$.%
\footnote{VDM ``mappings'' $m \in \mapof{D}{R}$ are finite constructed functions that can be thought of as sets of pairs from $D \x R$.
Use is made below of a domain operator ($\dom{m}$) that yields the current set of domain elements.}
These records have the following fields:

\begin{record}{JobInfo}
type: JobType\\
d: ClockValue\\
e: Duration
\end{record}

\noindent
Where the $type$ field is set to the contents of the static $JobType$;
the $d$ field is set to the actual deadline which is computed by adding the value in the $D$ field of $JobType$
to the time at which the job starts
(the current value of the $t$ field of the $State$);
the $e$ field is used to record 
the amount of resource (time) that the job has consumed.%
\footnote{This value is at least useful as a ghost variable in the specifications below;
it  is only actually necessary to maintain this value at run time for some scheduling algorithms.}



The third field of $State$ records which job
(if any) is actually running;
a scheduler sets $run$ to a $JobId$ that is in the domain of $active$ to make the corresponding job execute.%
\footnote{Whether one or more jobs can run at the same time depends on the number of processors;
here, only the single processor case is handled but it is straightforward to change $run$ so that it contains a (possibly empty) set of $JobId$s.}

Jobs themselves perform actions that affect shared entities ($shared-ents$ will at least contain variables). 

The $mode$ field of $State$ is explained and used in the discussion of  fault tolerance in Sect.~\ref{S-FaultT};
for now it is assumed that $mode = \const{Normal}$.



Rely and guarantee conditions are stated formally below 
but it is worth relating the general ideas of these interference predicates to the objects defined here.
A key planning assumption about execution of a job is that it will consume no more execution time ($e$)
than its estimated worst-case execution time ($C$ of $JobType$).
This becomes a guarantee condition for each $Job$;
the $Scheduler$ relies on this being the case for all jobs in $active$.

VDM allows for types to be restricted by (data type) invariants;
the second conjunct of the following invariant expresses a restriction about $active$ and $run$:%

\begin{fn}{inv-State}{st}\\
\signature{State \to \Bool}
st.mode = \const{Normal} \And \\
(st.active = \emptymap \And st.run = \nil \Or
st.run \in \dom{st.active}) \And\\
\forall{j \in \dom{st.active}}{st.t  \leq st.active(j).d} 
\end{fn}

\noindent
The final conjunct of the $inv-State$ predicate records the key scheduling condition
of any real-time system
that no job should execute beyond its deadline (i.e. that it must complete no later than its deadline).
It is important to remember that objects are considered to satisfy their invariants:
not only does $st \in State$ imply that $inv-State(st)$ holds
but any operation that manipulates objects of $State$ must preserve the invariant.
Invariants thus limit the states that satisfy rely and guarantee
(as well as pre and post)
conditions.



\subsection{$Time$ vs computer clocks}
\label{S-rho}



Deadlines and estimated execution times relate to the commonly accepted notion of time in the physical world;
software cannot detect this directly and can only read the internal clock of the hardware on which it runs.
To formalise this, 
the discussion here distinguishes $ClockValue$s from $Time$.
(Typically, $\alpha$ with or without subscripts is used as a value in the set $Time$.)
The overall set of objects used in the description of scheduling are mathematical functions from a dense $Time$ set to the $State$s 
defined in Sect.~\ref{S-State}:

\type{\Sigma}{Time \to State}

\noindent
Clearly,
little can be achieved unless the $ClockValue$s approximate $Time$ and here a notion from the research on 
time bands~\cite{BurnsHayes09} is used,
$=\sb{\rho}$ indicates that two values are equal to within the precision, $\rho$,  of the time band under discussion. %

The relationship between $\alpha \in Time$ and $t \in ClockValue$ is governed by $\mathcal{T}$:%
\footnote{Notice that $\mathcal{T/E/A}$
are mathematical functions rather than the VDM ``mappings'' used above.}

\begin{fn}{\mathcal{T}}{\sigma}\\
\signature{\Sigma \to \Bool}
  (\forall{\alpha \in Time}{\sigma(\alpha). t = \sb{\rho} \alpha}) \And\\
  (\forall{\alpha\sb{1}, \alpha\sb{2} \in Time}{\alpha\sb{1} < \alpha\sb{2} \Implies \sigma(\alpha\sb{1}).t \leq \sigma(\alpha\sb{2}).t})
\end{fn}

\noindent
The second conjunct of $\mathcal{T}$ ensures that the computer clock cannot go backwards!
This is necessary because the precision allows latitude:
for two sufficiently close (less than $\rho$) points in $Time$, their $State$s might be identical;
for two times more than $\rho$ apart, 
the first conjunct of $\mathcal{T}$ requires that the $t$ fields of their $State$s advance.

An execution-time scheduler can only make $e$ advance by moving the appropriate $JobId$ to $run$.
When a job is running,
its execution time is advanced in accordance with:%
\footnote{
If the run-time system has more than one processor and their speeds differ,
an abstract notion such as amount of work would have to be related to the differing processing speeds of the processors.}

\begin{fn}{\mathcal{E}}{\sigma}\\
\signature{\Sigma \to \Bool}
  \forall*{\alpha\sb{1}, \alpha\sb{2} \in Time}{
   \forall*{j \in (\dom{\sigma(\alpha\sb{1}).active} \inter \dom{\sigma(\alpha\sb{2}).active})}{
	((\forall{\alpha \mid \alpha\sb{1} \leq \alpha \leq \alpha\sb{2}}{j = \sigma(\alpha).run}) \Implies \T4
 			\sigma(\alpha\sb{2})(j).e \minus \sigma(\alpha\sb{1})(j).e =\sb{\rho} \alpha\sb{2} \minus \alpha\sb{1}) \And\\
	((\forall{\alpha \mid \alpha\sb{1} \leq \alpha \leq \alpha\sb{2}}{j \neq \sigma(\alpha).run}) \Implies \T4
 			\sigma(\alpha\sb{2})(j).e = \sigma(\alpha\sb{1})(j).e) 
	}
	}
\end{fn}

So,
using $\mathcal{A}$ from Sect.~\ref{S-f-planning},
the definition of $\Sigma$ with its invariant is:

\type{\Sigma}{Time \to State}
\where
\begin{fn}{inv-\Sigma}{\sigma}
\signature{\Sigma \to \Bool}
 \mathcal{A}(\sigma) \And \mathcal{T}(\sigma) \And \mathcal{E}(\sigma)
\end{fn}

\subsection{Specifying progress}
\label{S-time}



As indicated in Fig.~\ref{F-RG},
guarantee conditions limit what happens when the specified program makes a state change.
Specifying that progress must occur was not part of the initial rely-guarantee approach.
But it is of the essence of a scheduler that it must ensure progress of jobs 
and this would traditionally be viewed as a liveness condition
and perhaps be specified using some form of temporal logic.
However,
the model of $State$ used above,
together with the $\mathcal{T}$ part of the invariant on $\Sigma$,
forces $t$ to advance with respect to $Passage-of-Time$ and this opens up an alternative way to require progress.
By including time in the state values,
what are normally considered to be liveness conditions can be handled via invariants. %
The overall scheduling requirement on $State$ is expressed in the third part of the invariant shown above on $State$:

\begin{fn}{inv-State}{st}\\
\cdots \And\\
\forall{j \in \dom{st.active}}{st.t  \leq st.active(j).d} 
\end{fn}


\noindent
Recalling the discussion in Sect.~\ref{S-State} about invariants,
there is an obligation on $Scheduler$ to maintain $inv-State$
(i.e.~that all jobs terminate by their deadline);%
\footnote{This is in addition to the specified $guar-Scheduler$.}
it is a consequence of the frames
(which variables can be accessed by which operations)
that only the scheduler can maintain this invariant and that it can only do so indirectly:
when the scheduler sets the state variable
$run$ to contain the appropriate $JobId$ it moves towards the completion of that job 
at which point it will be removed from $active$.



\section{Specifying EDF}
\label{S-EDF-spec}



There are several scheduling approaches that are claimed to meet the requirements set out in Sect.~\ref{S-f-Scheduler}:
\cite{BurnsWellings-16} provides informal specifications of 
``fixed priorities'',
``least laxity'' and
``earliest deadline first''.
The third of these (EDF) is selected for detailed specification here.
The notion of correctness for any of the scheduling approach actually applies to the combination of $Planning$ and the run-time $Scheduler$.
But there is an unusual twist in that this is not a simple sequential combination of two operations since the former actually generates 
(or selects) 
the latter.%

Only the $Scheduler$ can actually cause jobs to execute and meet their deadlines 
but the argument that a specific scheduling algorithm
--in this case EDF--
will cope with the expected job arrival pattern and estimated WCETs is the responsibility of the (human) scheduler.

\begin{theorem}\label{Th-EDF}
EDF works:
providing the assumptions are satisfied at run time,
a $Scheduler$ that ensures the $Job$ with the earliest deadline is made to $run$ 
(within $\rho\sb{s}$ of any job arrival or completion)
will ensure that all jobs complete no later than their deadlines plus $\rho\sb{s}$.
\end{theorem}

\paragraph{Argument}
The justification that EDF will cope with a specific arrival pattern and collection of estimated WCETs
obviously depends on specific
$\mathcal{A}$ and $D/C$ for each $JobType$;
in general this calculation involves identification of the most pessimistic ``critical instants'' and the most difficult arrival pattern
(including the deadlines being so ordered as to maximise job switching).
A proof of conditions under which EDF scheduling suffices is in \cite{LL73}
and is formalised in~\cite{cerqueira2016prosa}.

\vspace{8pt}
\noindent
If the characteristics of the application are such that the preconditions of the EDF scheduler are not
satisfied (e.g. the WCETs turn out to be too large) then the Planning phase is aborted and the
system thus never starts its execution.

\subsection{Invariant for EDF}
\label{S-EDF-inv}

EDF is a dynamic (run-time) scheduling approach that does what it says:
an actual deadline ($d$) is computed for each job on its arrival
and the scheduler ensures that the job with the earliest deadline is in $run$. 
This is expressed in the following invariant:



\begin{fn}{inv-EDF}{st}\\
\signature{State \to \Bool}
st.mode = \const{Normal} \And \\
(st.run = \nil \Or\\
\forall{j \in \dom{st.active}}{ st.active(st.run).d \leq st.active(j).d})
\end{fn}


It is important to note that Theorem~\ref{Th-EDF} confirms that $Planning$ establishes that
--under the given workload assumptions--
maintaining $inv-EDF$ will preserve $inv-State$.

When the scheduler is created,
$inv-EDF$ is trivially established because there would be no jobs in $active$.
A qualification about this invariant is that the scheduler experiences job arrivals and completions so the EDF requirement
is treated as a ``class invariant''
on $Scheduler$ which is maintained by the methods of the class:

\begin{itemize}

\item $Arrival$: selects a fresh $JobId$ and updates $active$ by pairing that identifier with the $JobInfo$; 

\item $Completion$: removes the job from the $active$ map.

\end{itemize}


\noindent
Not only must $inv-EDF$ be re-established by the end of either method call,
this must be achieved within the precision of the time band  $\rho\sb{s}$ 
(perhaps 100~microseconds)
that was assumed in planning.

\subsection{Class and methods for EDF}
\label{S-EDF-class}



The overall requirements on the scheduler are given first followed by the specifications of its two methods.
There is no post condition for $Scheduler$ because it is not meant to terminate.

\begin{op}[Scheduler]
\ext{\Rd t: ClockValue\\
       \Wr active: \mapof{JobId}{JobInfo}\\
       \Wr run: \Opt{JobId}}
\pre{active = \emptymap}       
\rely{\forall{j \in \dom{active'}}{active'(j).e \leq active'(j).type.C}
}
\guar{\forall{j \in (\dom{active} \inter \dom{active'})}{
active'(j).type = active(j).type  
}}
\end{op}

\noindent
Notice that the constraints are on the primed value of $active$ in the rely condition to mark that the property must hold after activity of any other thread.
Similarly,
the relation between before and after values of $active(j)$ in the guarantee condition 
specifies that the scheduler must not modify the $type$ field of any active job.
As has been made clear above,
all states must satisfy both $inv-State$ and $inv-EDF$.



The specifications for the two methods of $Scheduler$ are:
%

\begin{op}[Arrival]
\args{jt: JobType}
\ext{\Wr active: \mapof{JobId}{JobInfo}\\
       \Wr run: \Opt{JobId}\\
       \Rd t: ClockValue}
\post{\exists*{j \in (JobId \minus \dom{active})}{active' = active \union \map{j \mapsto mk-JobInfo(jt, jt.D + t, 0)}}
}
\end{op}

$Completion$ is triggered by a job when it finishes;
the main point of the operation is to re-establish $inv-EDF$ 
by ensuring that the $JobId$ of the job next closest to its deadline is moved to $run$.

\begin{op}[Completion]
\args{id: JobId}
\ext{\Wr active: \mapof{JobId}{JobInfo}\\
       \Wr run: \Opt{JobId}}
       \pre{id = run}
\post{active' = \set{id} \dsub active 
}
\end{op}


Turning to the obligations on jobs:

\begin{op}[Job]
\ext{\Rd t: ClockValue\\
\Rd job: JobInfo\\
\Wr shared-ents: \mapof{Id}{Value}}
\rely{ job'.type = job.type \And 
t' \leq job'.d}
\guar{job'.e \leq job.type.C}
\post{work(shared-ents, shared-ents')}
\end{op}





\noindent
Crucially, 
the guarantee condition obliges each job to stay within its estimated worst-case execution time 
(and the transitive closure requires this to be true on job termination).
The undefined $work$ predicate in the post condition of $Job$ is a reminder that each job has work to do
that would result in changes to the shared entities ---
but specifying the detailed function of individual job types is not of concern here.

\subsection{Properties of EDF}

In addition to Theorem~\ref{Th-EDF},
checking that the respective obligations of $Scheduler$ and $Job$s match follows from two straightforward lemmas.

\begin{lemm}
The assumptions in $rely-Job$ are satisfied.
\end{lemm}

\paragraph{Argument}The conjuncts of $rely-Job$ follow from $guar-Scheduler$ (which rules out changes to $type$)
and $inv-State$ which follows from $inv-EDF$ by Theorem~\ref{Th-EDF}.

\begin{lemm}
\label{L-curly-A}
The assumptions in $rely-Scheduler$ are satisfied.
\end{lemm}

\paragraph{Argument}The collection of active jobs obeying their estimated worst-case execution times (WCET) ensures $rely-Scheduler$:

\begin{formula}
\forall*{st, st' \in State}{
\begin{formbox}
  (\forall{j \in \dom{st'.active}}{
     guar-Job(job, st'.active(j)))\Implies} 
     \end{formbox}}  \R
       rely-Scheduler(st , st')
\end{formula}



\subsection{Implementation}



The verification of programs with respect to pre/post conditions is well understood; 
furthermore~\cite{Hoare71a} showed that such specifications could be used as the basis of a formal development process and
tool support for approaches such as {Event-B}~\cite{Abrial10} has greatly increased the practicality of such activities.
There is less experience with developments from specifications employing rely and guarantee conditions so a few points are highlighted in this section.

There is no post condition for the scheduler because it is expected to run indefinitely long.
The obligations on the scheduler include maintaining the state invariant ($inv-EDF$) at each step in its execution.
The way that an implementation maintains $inv-EDF$ is necessarily indirect
because the execution time ($e$) for each job is increased (via $\mathcal{E}$) with respect to $Passage-of-Time$.
It is possible to view the quantification over all $active$ jobs as a prompt to get rid of jobs!
Since $guar-Scheduler$ limits access to $active$ to creation and deletion of jobs
(i.e.~it is not allowed to change $JobInfo$ of running jobs),
this can only be achieved indirectly.
A job removes itself from $active$ when it has performed its function so the scheduler must ensure that jobs make progress.
Furthermore,
progress must be timely because of the requirement that,
for all active jobs ($j$),
$t \leq active(j).d$.
Facilitating job progress amounts to making it run so that its used execution time ($e$) advances
and $\mathcal{E}$ shows that this only happens when $run = j$.
Thus a key requirement for the $Scheduler$ implementation is that it executes frequently enough to preserve the invariant.

An obvious implementation strategy that would tackle satisfying the specification of the scheduler would be
to check deadlines each time a new job arrives or the running job completes.




Implementing jobs so that they satisfy their guarantee conditions is,
in a sense,
the reverse of the true obligation:
the input to $Planning$ has to be reliable estimates of their WCET.

\section{Fault-tolerance}
\label{S-FaultT}

Specifying the fault-tolerant behaviour of run-time scheduling is a key contribution of the current paper.
The approach is to question assumptions that are made for optimal  performance
and to record weaker assumptions that characterise less desirable but safe  behaviour.


A specification of a sequential program using pre and post conditions
is satisfied by an implementation that can do absolutely anything if its execution starts in a state that does not satisfy the pre condition;
the implementation would not only be accepted if it produced random answers; 
it need not even terminate.
As made clear above,
rely conditions are also assumptions that an implementer is invited to make about the deployment context of the created artefact:
if the rely condition is not satisfied by a state transition of the environment,
the implementation is free of any obligation to the post condition, the guarantee condition or any invariant.
One possibility is that the implementation does something useful
but it is far sounder engineering to specify 
--explicitly--
layers of fault tolerant behaviour.


The key idea here is to address fault tolerance by questioning each of the clauses of the (optimistic) assumptions.
The example of jobs overrunning their estimated WCET is handled in detail in the remainder of the current section;
handling exceptions to the expected job arrival pattern
($\mathcal{A}$)
is mentioned in Sect.~\ref{S-close}. 

When a specification contains layers of rely and guarantee conditions,
an implementation is required to respect all of the layers.
Previous examples of specifying layers of fault tolerant behaviours with nested rely and guarantee include~\cite{BurnsHayesJones-19}.

\subsection{Dependability terminology}
\label{fef}



Fault tolerance is a large topic;
the current paper employs the useful three-fold distinction made in~\cite{RandellEt04} between faults, errors and failures.
Furthermore,
the injunction to identify to which systems the concepts are being applied is followed
(a failure in one system can manifest itself as a fault in another).
Rather than repeat the general definitions in~\cite{RandellEt04} of the three terms,
their use in the current context can be sketched as follows.
 
If a job overruns its estimated WCET,
this constitutes a failure of the job
but it is not necessarily a failure of the overall system.
As far as the (run-time) scheduler is concerned,
a fault has occurred that it might be possible to tolerate.
The scheduler is in an error state that might not result in it failing to meet its specification;
for example,
there might be enough slack in the current load that the error state can be tolerated.

It is of course also possible that the situation cannot be recovered in that the job fault propagates to a failure of the scheduler to meet its specification
and some job(s) might not complete by their deadlines.
Much the same series of fault, error and failure instances can be discerned with job arrival patterns
that do not match $\mathcal{A}$.

Schedulers can be designed to handle error states in various ways but the objective is to avoid jobs
(at least safety-critical ones)
running past their deadlines.
Choices about how closely to monitor execution time govern how well fault-tolerance is achieved
--- see Section~\ref{S-exceed}.


There is however another dimension of the fault, error, failure triage.
When one system creates another,
a failure in the creating process can give rise to an error in the created object.
For example, 
a programming failure, that leaves a bug in a designed program, creates an error that may never give rise to a failure.
A similar chain can be seen in schedule planning which, 
if not done properly,
can result in a run-time scheduler that has errors that might result in failures
(for example, deadlines may be missed if the maximum allowed load on the system
does actually materialise).

\subsection{Exceeding WCET}
\label{S-exceed}



This sub-section reverts to the level of general run-time schedulers as in Sect.~\ref{S-f-Scheduler};
this makes it possible to outline how nested rely and guarantee conditions can be used to specify fault-tolerance;
the specifics of how EDF handles fault-tolerance are discussed in Sect.~\ref{S-EDF-exceed}.

If at some point in time,
the execution of a job fails by exceeding its estimated worst-case execution time:

\begin{formula}
\exists{j \in \dom{active}}{active(j).e > active(j).type.C}
\end{formula}

\noindent
 a non-fault-tolerant scheduler that just aborted at that time would be deemed to meet the specification in Sect.~\ref{S-f-Scheduler} ---
 but this is clearly not resilient.
 A simple attempt at a fault-tolerant reaction might be to drop
 (or background)
 any offending job and this approach can be used to illustrate the idea of layered rely-guarantee descriptions
 (but a more useful fault-tolerant defence is developed below).
 
 Dropping overrunning jobs could be specified using a weaker invariant (than that in Sect.~\ref{S-State}):
 
\begin{fn}{inv-State-W}{st}\\
\signature{State \to \Bool}
st.mode = \const{Normal} \And \\
(st.active = \emptymap \And st.run = \nil \Or
st.run \in \dom{st.active}) \And\\
\forall*{j \in \dom{st.active}}{st.active(j).e \leq st.active(j).type.C  \Implies st.t  \leq st.active(j).d} 
\end{fn}

Remembering that 
--along with $guar-Scheduler$--
preserving the invariant is an obligation on the scheduler,
this minimal change illustrates the idea of layers of specification:
with strong assumptions, 
the implementation must meet strong obligations;
where only weaker assumptions hold,
the implementation has weaker obligations to fulfil.

A trivial lemma shows that the new invariant is a weakened form of the original:

\begin{lemm}
$inv-State(st) \Implies inv-State-W(st)$
\end{lemm}

A valid implementation must meet both sets of requirements.
(The issue of returning to the more optimistic mode is discussed in Sect.~\ref{S-revert} with a more interesting degree of fault tolerance.)
  
 In an EDF implementation,
 there would a problem that an overrunning job could continue to absorb resources and force other jobs to miss their deadlines.
 The obvious solution is to remove overrunning jobs from $active$ but the next section describes a more nuanced fault-tolerant implementation.

\subsection{Levels of criticality}
\label{S-MC}
 
Mixed-Criticality Scheduling~\cite{Ves2007} enhances the robustness of schedulers 
by distinguishing different levels of ``criticality'' for classes of  jobs. 
A practical system might offer many levels of criticality 
but the approach to handling mixed criticality can be illustrated with two ($HI$-crit vs. $LO$-crit).
Obviously,
the intention is to give preference to $HI$-crit jobs 
and, 
when there is insufficient resource, 
to do this at the expense of $LO$-crit jobs.

Unfortunately,
it is insufficient to detect just before the deadline of a $HI$-crit job that a $LO$-crit job with an earlier deadline has used resources that could have been more wisely allocated to the job with higher criticality.
The MCS way to avoid this trap is to associate an early ``virtual deadline''~\cite{BBDLMMS2011b}  with any $HI$-crit job:
this is the value in the $D$ field of their $JobType$
and is calculated during planning.
An optional $HI-crit$ object contains an additional deadline allowance ($AD$) that, when
added to the job's current deadline, represents the actual deadline requirement of the application.
Furthermore,
an extra (more conservative) estimated WCET is included in $HC$ (with $HC > C$).
These two values are located in a $HI-crit$ object; 
the $X$ field of $JobType$ is $\nil$ for $LO$-crit jobs:

\begin{record}{JobType}
D: Duration\\
C: Duration\\
X: \Opt{HI-crit}
\end{record}

\begin{record}{HI-crit}
AD: Duration\\
HC: Duration
\end{record}

A useful predicate that simplifies some expressions below is:

\begin{fn}{is-hi}{job}
\signature{JobInfo \to \Bool}
job.type.X \neq \nil
\end{fn}



 The final $mode$ field of $State$ in Sect.~\ref{S-State} distinguishes between $\const{Normal}$ mode when all jobs are scheduled and
 $\const{Ft}$ mode during which $LO$-crit jobs are ignored
 (they could be terminated or just backgrounded until there is spare capacity).
So, 
in the fault-tolerant mode, 
$inv-State-FT$ only requires that all $HI$-crit jobs meet their deadlines:

\begin{fn}{inv-State-FT}{st}\\
\signature{State \to \Bool}
inv-State(st) \Or \\
st.mode =\const{Ft} \And \\
st.active \neq \emptymap \And st.run \in \dom{st.active} \And is-hi(st.active(st.run))  \And \\
\forall{j \in \dom{st.active}}{is-hi(st.active(j)) \Implies st.t  \leq st.active(j).d} 
\end{fn}

\noindent
Notice that the possibility of $active = \emptymap$ can be ignored because this would trigger return to normal mode
(see Sect.~\ref{S-revert}).

It is trivially true that:

\begin{lemm}
$inv-State(st) \Implies inv-State-FT(st)$
\end{lemm}

\subsection{EDF handling of exceeding WCET}
\label{S-EDF-exceed}

Turning to how EDF behaves in fault-tolerant mode:
as well as Theorem~\ref{Th-EDF} still applying to $\const{Normal}$ mode operation,
it must be true in $\const{Ft}$ mode that:

\begin{theorem}\label{Th-EDF-VD}
Fault-tolerant EDF works:
providing the arrival assumption ($\mathcal{A}$) is satisfied and the $HI$-crit jobs execute within their $HC$ estimated execution times,
a $Scheduler$ that ensures the $Job$ with the earliest deadline is made to $run$ 
(within $\rho\sb{s}$ of any job arrival or completion)
will ensure that the $HI$-crit jobs complete no later than their (real) deadlines plus $\rho\sb{s}$.
\end{theorem}

\paragraph{Argument}
As in  Theorem~\ref{Th-EDF},
the justification that EDF will cope with a specific arrival pattern and collection of extended WCETs
requires a calculation that involves identification of the most pessimistic ``critical instants'' and the most difficult arrival pattern.

The EDF invariant becomes:

\begin{fn}{inv-EDF-FT}{st}\\
\signature{State \to \Bool}
inv-EDF(st) \Or \\
st.mode =\const{Ft} \And \\
is-hi(st.active(st.run))  \And \\
\forall*{j \in \dom{st.active}}{is-hi(st.active(j)) \Implies 
st.active(st.run).d \leq st.active(j).d  
}
\end{fn}

\begin{lemm}
Thanks to the use of the virtual deadline,
it is again trivial that:

\begin{formula}
inv-EDF(st) \Implies inv-EDF-FT(st)
\end{formula}
\end{lemm}

%

In the fault-tolerant mode,
$guar-Job-FT$ only applies to $HI$-crit jobs and uses $HC$:

\begin{op}[Job-FT]
\ext{\Rd t: ClockValue\\
\Rd job: JobInfo\\
\Wr shared-ents: \mapof{Id}{Value}\\
\Rd mode: \set{\const{Normal},\const{Ft}}}
\rely{ job'.type = job.type \And 
t' \leq job'.d}
\guar{mode =\const{Ft} \And is-hi(job) \Implies job'.e \leq job.type.X.HC}
\post{work(shared-ents, shared-ents')}
\end{op}

The overall scheduler class becomes:

\begin{op}[Scheduler-FT]
\ext{\Rd t: ClockValue\\
       \Wr active: \mapof{JobId}{JobInfo}\\
       \Wr run: \Opt{JobId}\\       
       \Wr mode: \set{\const{Normal},\const{Ft}}}
\pre{active = \emptymap}       
\rely{mode = \const{Normal} \And \\
           (\forall{j \in \dom{active'}}{
                   active'(j).e \leq active'(j).type.C) \Or}\\
mode =\const{Ft} \And \\
           (\forall*{j \in \dom{active'}}{
                    is-hi(active(j)) \Implies active'(j).e \leq active'(j).type.X.HC)}
}
\guar{\forall{j \in (\dom{active} \inter \dom{active'})}{
active'(j).type = active(j).type  
}}
\end{op}

Moving on to the methods of the $Scheduler-FT$ class.
The $Arrival-FT$ method in $\const{Normal}$ mode is as above;
in $\const{Ft}$ mode any arriving $LO$-crit jobs are ignored and the deadlines for $HI$-crit jobs are set to the extra deadline.

\begin{op}[Arrival-FT]
\args{jt: JobType}
\ext{\Wr active: \mapof{JobId}{JobInfo}\\
       \Wr run: \Opt{JobId}\\
       \Rd t: ClockValue\\
       \Rd mode: \set{\const{Normal},\const{Ft}}}
\post{
  mode = \const{Normal} \And \T2
         (\exists*{j \in (JobId \minus \dom{active})}{active' = active \union \map{j \mapsto mk-JobInfo(jt, jt.D + t, 0)}) \Or} \\
  mode =\const{Ft} \And \T2
    (jt.X = \nil \And active' = active \Or \T2
   jt.X \neq \nil\And \T2
    \exists*{j \in (JobId \minus \dom{active})}{active' = \R
          active \union \map{j \mapsto mk-JobInfo(jt, (jt.D + jt.X.AD + t), 0)})}
}
\end{op}

The $Completion$ method is unchanged
(but could trigger $Mode-up$ when the final job is removed from $active$).

If a $HI$-crit job is in trouble,
an EDF-FT implementation has to switch to $\const{Ft}$ mode.
There are several ways in which an implementation could detect trouble:
a last resort is to wait until a $HI$-crit job hits its (virtual) deadline.
Here,
it is assumed that the mode switch occurs when a $HI$-crit job overruns its optimistic estimate of WCET.
%
%
%
%


This could be triggered by adding to $inv-EDF-FT$ a conjunct that:

\begin{formula}
st.mode = \const{Normal} \Implies\R
\forall{j \in \dom{st.active}}{st.active(j).e \leq st.active(j).type.C}
\end{formula}

\noindent
This can be achieved either by removing $LO$-crit jobs from $active$ or by switching modes as follows:
a new method is required for the $Scheduler$;
this handles mode change from normal to $\const{Ft}$ mode:%
\footnote{The implicit definitions of mappings in the post condition are a simple extension of implicit set specifications 
such as
$\set{i \mid is-prime(i)}$.}

\begin{op}[Mode-down]
\ext{\Wr active: \mapof{JobId}{JobInfo}\\
        \Wr mode: \set{\const{Normal},\const{Ft}}}
\pre{mode = \const{Normal} \And\\
			\exists{j \in \dom{active}}{is-hi(active(j)) \And active(j).e > active(j).type.C}}
\post{mode' =\const{Ft} \And\\
         active' = \T2
			\map{j \mapsto active(j) \mid j \in \dom{active} \And \Not is-hi(active(j))} \union \T2
			 \map{j \mapsto mk-JobInfo(jt, (d + jt.X.AD), e) \mid \R
			 		j \in \dom{active} \And 
					is-hi(active(j)) \And \R
					active(j) = mk-JobInfo(jt, d, e)} 
}
\end{op}


\noindent
The post condition describes the most commonly adopted option of setting all $HI$-crit jobs to use their extended deadlines and
leaving the $LO$-crit jobs in $active$
(but the invariant leaves the freedom for them to be ignored);
describing other options is straightforward.


%

%

It is worth repeating that a valid implementation of a layered rely-guarantee specification must meet all of the layers.
Thus, a fault-tolerant EDF-FT implementation is expected to complete all jobs
--whether $HI$-crit or $LO$-crit--
by their deadlines if the optimistic assumptions hold.

\subsection{Reverting to normal mode}
\label{S-revert}

Having specified what should happen in one of the situations in which a fault-tolerant EDF system
recovers from a failure of a job to stay within its (optimistic) WCET,
the question naturally arises as to how to return to normal 
(or optimistic) mode.
One example of a safe situation is where there are no jobs executing:

\begin{op}[Mode-up]
\ext{\Rd active: \mapof{JobId}{JobInfo}\\
        \Wr mode: \set{\const{Normal},\const{Ft}}}
\pre{active = \emptymap}
\post{mode' = \const{Normal}
}
\end{op}

There are weaker conditions such as only $LO$-crit jobs remaining but this one illustrates how $Mode-up$ might be specified.


\section{Closing}
\label{S-close}

\subsection{Related results}


Another place for fault tolerance can be identified by 
questioning what needs to be done when the assumption about arrival times ($\mathcal{A}$) is violated:
its treatment requires more data to be kept in the run-time state but would follow similar approach to that in Sect.~\ref{S-exceed}.

As indicated above,
there are alternatives to the EDF approach to scheduling
(see~\cite{BurnsWellings-16}).
If the planning stage selects Fixed Priority scheduling, 
it is not necessary to compute the actual deadline ($d$) at run-time.
Another alternative is to arrange that the run-time scheduler ensures that the job with least laxity until its deadline is the one chosen for execution.
Although not written out in detail,
the approach to analysing these approaches appears to be similar to the treatment of EDF in Sect.~\ref{S-FaultT}.
 
%

There are of course many other details of scheduling that could be investigated using the models and proposed specification approach.
For example with jobs sharing (i.e.~concurrently accessing)
resources that must be accessed under mutual exclusion,
various resource sharing protocols exist such as
allowing $d$ to be less than $t+D$ for short periods of time.
Also one of the locking mechanisms for shared variables, 
called the ``deadline floor protocol''~\cite{10.1109/TC.2014.2322619}
would need to change $d$.

\subsection{Relation to earlier papers}

The current paper continues the line of research reported in~\cite{BurnsJones:ECRTS:2022,JonesBurns-22} 
that seeks to get clear specifications of fault-tolerant real-time scheduling;
in particular there is an emphasis on identifying assumptions made in the designs.
The conference paper~\cite{BurnsJones:ECRTS:2022} focussed on the scheduling issues
whereas~\cite{JonesBurns-22} contains the first use of the $\Sigma/State$ distinction to pin down the distinction between 
the sets $Time$ and $ClockValue$;
both papers also had more discussion than here on timeband ideas.%
\footnote{Not only is the current paper an extension of the earlier papers
--most notably with respect to analysing the fault tolerance--
even large parts of the current text have been rewritten since the initial submission in response to legitimate complaints from a referee 
that the order of presentation led to too many forward references.}


It is perhaps worth reviewing alternatives to some of the choices embodied in the current paper.
The limited use of timeband concepts~\cite{BurnsHayes09} in the current version should not be taken as our retreating from their importance.
Whilst it is true that the notion of $t =\sb{\rho} \alpha$ could be replaced by $t \in \set{\beta \mid \alpha \minus \rho/2 \leq \beta \leq \alpha + \rho/2}$,
this perpetuates the mistake of mapping all times down to the finest time axis.
Hopefully,
the reader will see that at least discussing the $Passage-of-time$ during planning is better conducted at a much coarser granularity.
The current authors also prefer to think about notions of granularity such as $=\sb{\rho}$ as approximations at a chosen granularity.

Turning to the use of rely and guarantee conditions, 
it is hopefully clear that standard post conditions that are relations between notions such as $State$ cannot cope with concurrent changes.
It would, of course, be possible to add ``ghost variables'' to the basic states and express interaction in terms of histories of values.
The current authors find that reasoning about interference in terms of rely and guarantee conditions far preferable.

A final decision that might be questioned is that this paper eschews the use of any Temporal Logic.
As well as the arguments for rely-guarantee conditions in the previous paragraph,
it is worth remembering that so-called ``Temporal Logics'' do not deal with time in the sense of $Time$ as used in Sect.~\ref{S-rho}.

\subsection{Related work}
\label{S-related}


The research reported in~\cite{cerqueira2016prosa,PROSA-2} is both supportive of the endeavours in the current paper and interestingly different.
For example~\cite{cerqueira2016prosa} starts with the rallying call:

\begin{quotation}
\noindent
For a field that prides itself in the development and analysis techniques for safety-critical systems---famously {\em hard} 
real-time systems---there recently has been an alarming number of errata correcting or retracting unsound schedulability analyses.
\end{quotation}
Their paper goes on to describe a framework for proofs that are mechanically checked  using Coq.
The authors emphasise,
moreover,
the need for readability of the proofs to ensure that they achieve what the user intended.
This work is clearly relevant to our aims.
Their target is to prove  conditions under which schedule planning will work
whereas we have worked on the specification of run-time schedulers ---
in particular on fault-tolerant aspects.

%

With respect to other differences between the two approaches~\cite{cerqueira2016prosa}  contains further
examples including their handling of multi-processor implementations and
their proof that, 
if any schedule exists,
an EDF schedule will work.

In a further quote from~\cite{cerqueira2016prosa}, they write:

\begin{quotation}
\noindent
In fact,
it is relatively easy to (accidentally) build formal specifications and proofs that are incomprehensible to non-experts.
\end{quotation}

\noindent
We would  echo this point and make clear that we are at the first of these stages:
we wish to look at the overall scheduling objective and to contribute understandable {\em specifications}.
One example is the distinction made above between planning and the run-time scheduler.
Furthermore,
the emphasis in the current paper on recording assumptions using rely and guarantee conditions
offers a systematic way of investigating fault tolerance by considering what actions can be added to the basic algorithms 
to limit the impact of violations of the assumptions.

Having looked at some of the {\sc Prosa} proof scripts,
there is clearly considerable scope for collaboration which the current authors intend to pursue once this paper is finalised.

The recent~\cite{VanhemsEt-22} also reports on use of Coq as an interactive theorem proving assistant.
In contrast to our approach,
they start at the level of a program and lift their results towards an overall specification.
Their main result is to address (complete with available proof scripts)
the optimality of EDF.

Both~\cite{cerqueira2016prosa} 
and~\cite{VanhemsEt-22}
are valuable contributions and there is clearly scope for potential collaboration.

\subsection{Concluding remarks}

The objective in writing this paper was to employ existing ideas on the rely-guarantee approach
and extend them where necessary so that they can be used to specify the temporal properties of real-time systems, and more generally
Cyber-Physical systems with their need for fault-tolerant behaviour and their support for Mixed-Critically (MC) applications.
To achieve this it has been necessary to define a relationship between the external `passage of time' and an
internal access to `clock time'. This relationship is then used to define invariants and rely conditions that force
the run-time scheduler to progress the application's jobs so that they will meet their deadlines. 

Fault tolerance
is addressed by challenging these invariants and forcing the system to transition to a degraded mode which is
defined by weaker assumptions and weaker guarantees. The choice of which scheduling approach to employ
and the application of the associated schedulability analysis is undertaken in a distinct Planning phase. In
this paper the EDF scheme is chosen as an illustration and the EDF-VD extension for fault tolerance and MC 
behaviour is applied.
This leads to a formal specification of a run-time scheduler and a precise definition of the assumptions applied
during planning.

\subsection*{Acknowledgements}

The authors are grateful for the detailed reports from referees on the initial submission.

The EPSRC ``Strata'' Platform Grant funded the initial collaboration of the current authors.
Jones' research is supported by a grant (RPG-2019-020) from the Leverhulme Trust.

\bibliography{master}

\end{document}